# Gaseous Detector with sub keV Threshold
# to Study Neutrino Scattering at Low Recoil Energies

A.V.Kopylov[*], I.V.Orekhov, V.V.Petukhov, A.E.Solomatin

*Institute of Nuclear Research of Russian Academy of Sciences*

*117312 Moscow, Prospect of 60[th] Anniversary of October Revolution 7A*

### Abstract

Gaseous detector with a sub keV electron equivalent threshold is a very perspective tool for the precision measurement of the neutrino magnetic moment and to observe coherent scattering of neutrinos on nuclei. The progress in the development of low noise electronics makes it possible to register the rare events at the threshold less than 100 eV. The construction of the gaseous detector is given and the typical pulses with amplitudes of a few eV observed on a bench scale installation are presented. The possible implications for future experiments are discussed.

### Introduction.

First neutrino detector made by F.Reines and C.L.Cowan 60 years ago opened a new field of a fascinating research with neutrino which seemed to be nearly undetectable object, as it was noted by W.Pauli himself. Since that time tremendous success in this field has been achieved. But till now neutrino continues intriguing the researchers and promises new discoveries. Development of the detectors with a threshold below 1 keV electron equivalent for experiments with reactor antineutrinos may, indeed, lead to new discoveries in neutrino physics. So far the lowest threshold achieved in these experiments is 3 keV, see, for example, GEMMA experiment [1] where the upper limit for the neutrino magnetic moment $2.9 \cdot 10^{-11}$ $\mu_B$ has been set. For the Dirac neutrino magnetic moment with standard model interactions $\mu_\nu \sim 3 \cdot 10^{-19}$ $\mu_B$ ($m_\nu$/1eV) which is far below the sensitivity of the experiment. However, in the extension of MSSM which, for example, includes a vectorlike leptonic generation which contains a fourth leptonic generation along with its mirrors a magnetic moment for the neutrinos as large as $10^{-12}$ $\mu_B$ can be obtained [2]. This is a basic motivation for experimentalists to refine the technique of its measurements. In the dark matter experiment CoGeNT [3] the germanium detector has been developed with a threshold of about 0.3 keV electron equivalent. Any further substantial reduction of the threshold for the semiconducting devices would be hardly possible. To progress in this field of research one highly demands the development of the detectors with a threshold of about 100 eV electron equivalent and may be even lower. This is a crucial point especially for the discovery of a coherent scattering of neutrinos on atomic nuclei (CNNS). The process of neutral current neutrino-nucleus elastic scattering has been described long ago [4, 5] but till now it remains to be a challenge for experimentalists. In CNNS the amplitude is a superposition of the individual amplitudes of scattering off each nucleon with relative phase factors.

---

[*] e-mail: beril@inr.ru

$$\left(\frac{d\sigma}{dT}\right)_c = \frac{G_F{}^2 M_A}{2\pi}\left(2 - \frac{2T}{E_\nu} + \left(\frac{T}{E_\nu}\right)^2 - \frac{M_A T}{E_\nu{}^2}\right)\frac{Q_W{}^2}{4}F^2\left(Q^2\right) \tag{1}$$

where $Q_W = N - Z\left(1 - 4\sin^2\theta_W\right)$ is a weak charge of a nucleus, $G_F$ is the Fermi coupling constant, $M_A$ is the mass of the nucleus, $F^2(Q^2)$ is the nuclear form factor [6]. Due to a coherence the cross-section is proportional to the square of a number of nucleons and may be so high that even a detector with 1 kg of a target material placed near the core of a commercial nuclear reactor may have a count rate from CNNS on the level of 1000 events per year for the flux of reactor antineutrinos of about $10^{13}$ ν/cm$^2$/s. In the GEMMA experiment at the Kalinin Nuclear Power Plant (KNPP), where a high-purity germanium detector of 1.5 kg has been placed 13.9 m from the 3 GW reactor core, the flux of antineutrinos was $2.7 \cdot 10^{13}$ ν/cm$^2$/s. Just for comparison, to get a similar count rate on the beam of neutrinos from a spallation neutron source [7] one needs a detector of a few tons scale. In Fig.1 it is shown the contribution of the effect from magnetic moment of neutrino to the total cross section of νe$^-$ scattering [8]

$$\frac{d\sigma}{dT} = \left(\frac{d\sigma}{dT}\right)_W + \left(\frac{d\sigma}{dT}\right)_\mu \tag{2}$$

$$\left(\frac{d\sigma}{dT}\right)_W = \frac{2G_F m_e}{\pi}\left[g_R{}^2 + g_L{}^2\left(1 - \frac{T}{E_\nu}\right) - g_R g_L \frac{m_e T}{E_\nu{}^2}\right] \tag{2a}$$

$$\left(\frac{d\sigma}{dT}\right)_\mu = \frac{\pi\alpha^2}{m_e{}^2}\left(\frac{1}{T} - \frac{1}{E_\nu}\right)\left(\frac{\mu_\nu}{\mu_B}\right)^2 \tag{2b}$$

where $g_L = \sin^2\theta_W$, $g_R = g_L + 1$

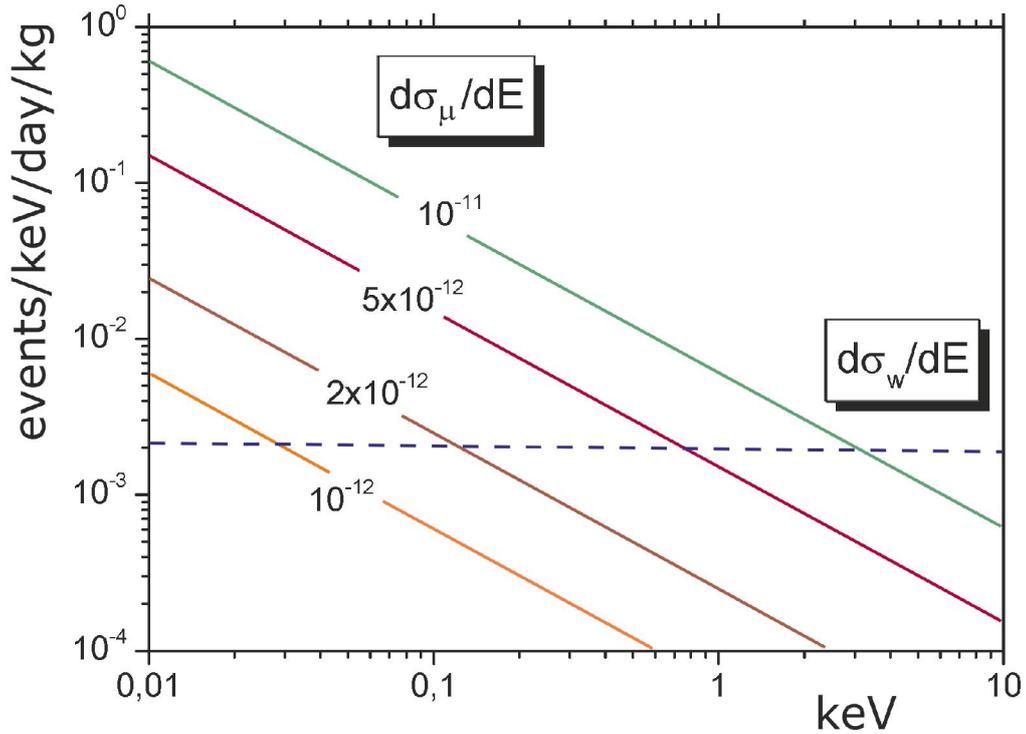

Figure 1. The count rate for antineutrino-electron elastic scattering as a function of energy in keV for different values of magnetic moment. By dashed line is indicated the count rate by standard electroweak model. The flux of antineutrinos is taken here $10^{13}$ ν/cm²/s.

One can see that to observe the effect from magnetic moment less than $10^{-11}$ μB the detector should have a threshold below 0.1 keV electron equivalent. With a threshold of a few tens of eV it will be possible to reach a limit $2 \cdot 10^{-12}$ μB. The notable feature of this scattering is that the cross section does not depend on the target because it is a pure νe⁻ elastic scattering. The situation is drastically different at coherent scattering of neutrinos on nuclei. CNNS can be observed only at small recoil energies when neutrino scatters on a nucleus as a wave on a grid, here the "grid" is composed of the nucleons tied by strong force with the rest of a nucleus. The energy of the recoiled nucleus depends on a mass of a nucleus, the heavier nucleus the smaller the energy of the recoils. This constitutes the "signature" of this process and may be used for events identification to determine what is responsible for the effect: νe⁻ elastic scattering or coherent scattering. Thus it would be useful to make measurements with different targets. High cross section makes it very attractive for experimentalists but low (less than 1 keV electron equivalent) recoil energy of a nucleus makes it very difficult for practical implementation. In Fig.2 it is shown the simulated effect for different nuclei.

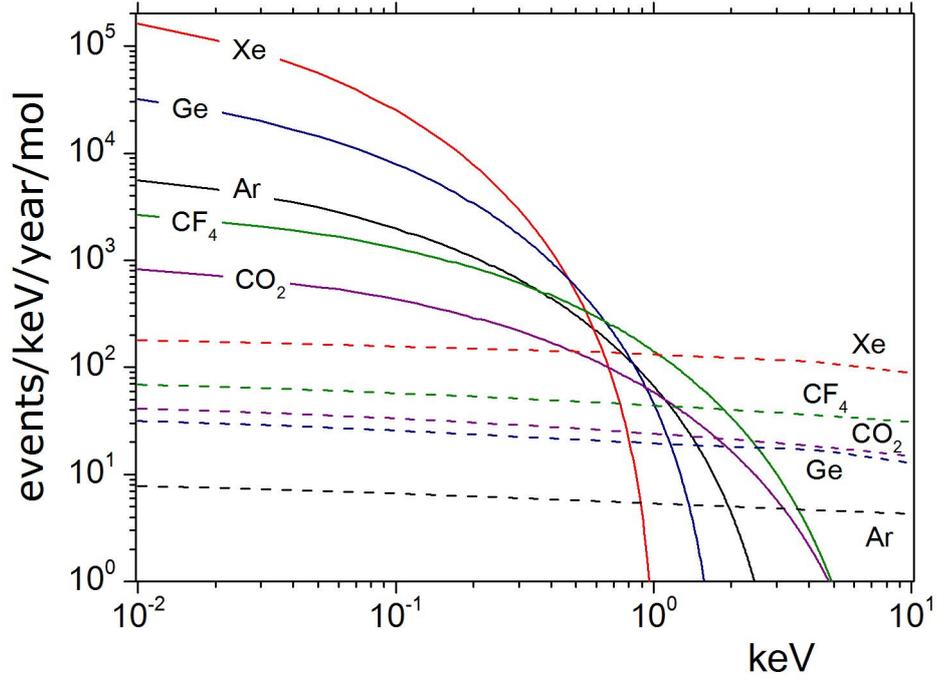

Figure 2. The count rate from CNNS as a function of energy of the recoiled nucleus for different gases. By dashed lines are indicated the count rates from neutrons generated by cosmic rays. The flux of antineutrinos is taken here $10^{13}$ v/cm$^2$/s.

One can see that for heavy nuclei the cross section is higher, but the energy of the recoils of the nuclei is smaller. This demands a detector with a very low threshold. The heavier nuclei, the more severe are these demands. By dashed lines is indicated the background from neutrons generated by cosmic rays at the site of KNPP. It was shown in [1] that background from fast neutrons of the reactor by adequate shielding is lower than from neutrons generated by cosmic rays. The precise (with the uncertainty of about 1-2%) measurement of CNNS will be useful for the study of neutron form factor $F^2(Q^2)$ of a nucleus and may be even $\sin^2\theta_W$ presented in a weak charge of a nucleus (see expression (1)). What is also crucial is the possibility to verify the discovery of this process by an independent way. Thus it is very important to develop not one, but several techniques to search for CNNS. In Fig 3 it is shown the rates for xenon (a) and argon (b) from coherent scattering on nuclei and from ve$^-$ scattering.

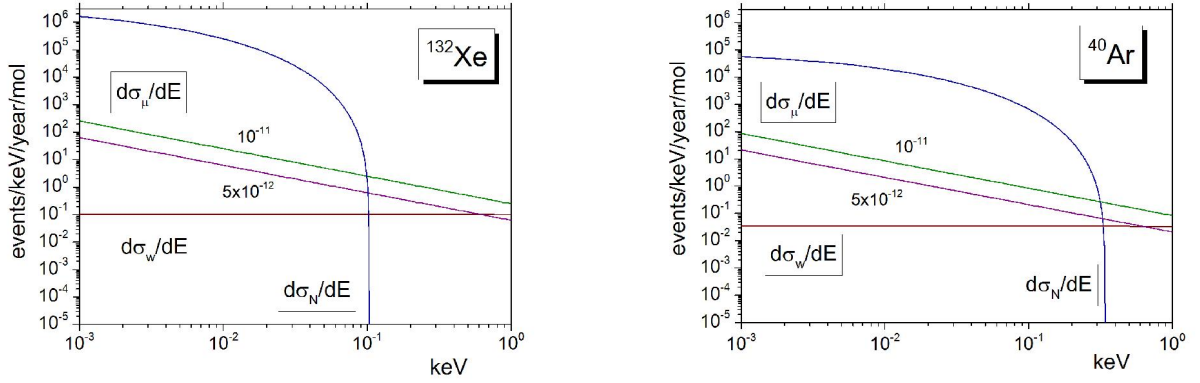

Figure 3. The count rates in xenon (left) and argon (right) for coherent scattering on nuclei and from $\nu e^-$ scattering at magnetic moment $5 \cdot 10^{-12}$ $\mu_B$ and $10^{-11}$ $\mu_B$.

Here we should take into account that only a small fraction of the nuclear recoil energy is used in the ionizing process. This fraction is called a quenching factor (QF) and for the recoil energy of the nuclei less than a few hundreds of eV it may be very different for different gases. To measure this factor for some specific gas used in experiment is a very crucial point and actually not an easy task. It demands the special efforts which may be very rewarding because it may give a very interesting result. Here is taken as a very probable value $QF = 0.1$ for recoiled nuclei. One can see that the effect from coherent scattering is absolutely predominant in the lower range of energy. The result of this is that for xenon it would be possible to extract the effect from magnetic moment if it is higher or equal to $5 \cdot 10^{-12}$ $\mu_B$ while for argon the lower limit is somewhere around $10^{-11}$ $\mu_B$. If $QF < 0.1$ this will shift the spectrum from coherent scattering to lower energies and this will be more appropriate for registration of the effect from magnetic moment of neutrino. From this point of view the most perspective gas to search for magnetic moment would be xenon. It has also advantage of having high $Z$ what is very important for measuring the effect from $\nu e^-$ elastic scattering. If the gas amplification can be as high as $10^5$ the threshold can be as low as 10 eV and then xenon is also good for detection of coherent scattering. The very promising aspect is also the possibility to use isotopically enriched xenon which will enable to reduce the background from internal radioactivity. At present several groups are working in this field. Both antineutrinos generated at nuclear reactors with the energy of several MeV and neutrinos of higher energies (of several tens of MeV) produced in the decays of pions and muons in the high intensity proton beams are planned to be used. For the first ones the mass of a target material should be of several kilograms, for the second ones – several tons to see the effect from CNNS. The recoil of a nucleus for the first case is expected to be less than 1 keV, for the second case – the tens of keV. Obviously, each approach has strong and weak points. Here we discuss the possibility to use a special construction of a gaseous detector of ionizing eradiation as a detector of CNNS. By choosing the gaseous proportional counter the emphasis is done on the following advantages of this technique:

1. Very high factor of the gas amplification ($> 10^4$).

2. Possibility to use gas at relatively high pressure about 1 MPa to obtain the mass sufficient for count rate of about 1 event per day.

3. Good signature of the events by a pulse shape (very characteristic front and tail of the pulses).

4. The possibility to discriminate noise from electromagnetic disturbances and microphonic effect.

5. Availability of the efficient methods of gas purification.

6. Detector can be fabricated only from very pure materials without PMTs as a possible source of ionizing eradiation etc.

7. The possibility to change easily the working gas ($CO_2 - CF_4 - argon - xenon$) not changing the configuration, what is important to perform the comparative measurements at the same site. The crucial point will be the background observed at very low recoil energy of nucleus. This is the main factor limiting the accuracy which can be achieved in experiment.

**Materials and Methods.**

1. The design of the detector.

The count rate from CNNS of reactor antineutrinos is predicted to be about 100 events per year per kg of a target material in the energy range from 20 eV to 200 eV of produced ionization. To collect a mass of about 1 kg in gas phase one needs a detector with a volume of about 50 liters even at pressure of about 1 MPa. But to get the gas amplification higher $10^4$ at High Voltage 3 kV the diameter of the cathode should be 40 mm, not more. To reconcile these conflicting demands we should use an array of counters and each one apart of a central, avalanche region with a small diameter of a cathode should have also external, drift region, separated from avalanche region by a grid. The diameter of the drift region is taken to be 140 mm. Apart from this, there should be external cylindrical layer of counters working as an active shielding and also as a passive one of the fluorescence from the walls of the counter. All assembly is placed in a cylindrical body made of titanium as a relatively pure on $^{226}$Ra material, as our previous measurements have shown [9]. In Fig.4 we show the general view of this counter.

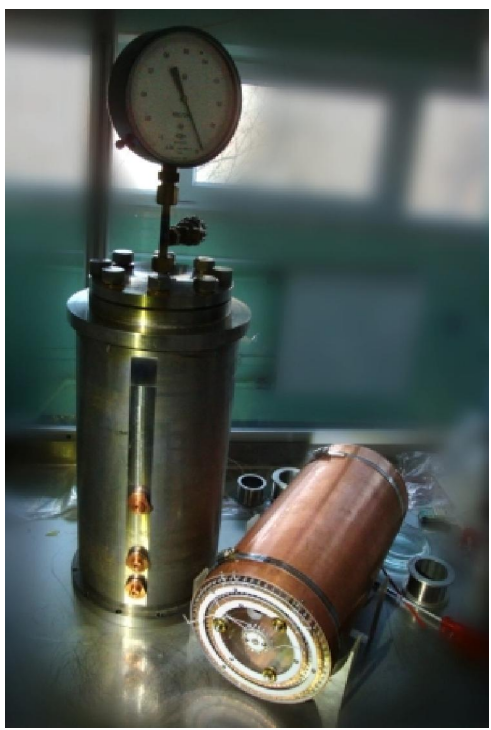

Figure 4. The counter before assembling.

We plan to use an array of 16 similar counters, each working on separate charge sensitive preamplifier and digitizing board. The counters will be assembled in 4 planes, each one having 4 counters. The size of the assembly will be approximately 100x100x100 cm. To reduce the background from cosmic rays, neutrons and gamma-rays the assembly will be placed in the box made of slabs of iron 30 cm thick, internal surfaces will be lined by borated polyethylene 40 cm thick. To shield from fast neutrons from the reactor we plan to use additional external layer of water 50 cm thick and on the outside - plastic scintillator as an active veto shield from ionizing particles of cosmic rays penetrating to the depth of about 16 m of water equivalent. The water shield reduces the background from fast neutrons by an order of magnitude, thus, it will be possible by comparing the data collected with and without water to prove that the contribution of reactor neutrons to the effect is negligible. All this assembly will be placed in a hermetically sealed housing filled by argon purified of radon. We select this design of shielding to reduce at most the background from gamma-quanta from external radioactivity and from neutrons, generated in iron by cosmic rays. Borated polyethylene 40 cm thick decreases approximately 100-fold the flux of fast neutrons from iron. The slabs of iron 30 cm thick effectively absorb gamma-radiation from the walls. In Fig.2 we show the calculated effect from CNNS and the background from neutrons, generated by cosmic rays at 16 meters of water equivalent for different working mediums of the detector.

## 2. The technique of the pulse shape discrimination.

We performed the measurements of the energy spectra of the pulses in argon using a small bench scale assembly. Proportional counter had 37 mm in diameter, the central wire of 20 μm and it was filled by argon and methane (10%) mixture by 100 and 300 kPa. A schematic of the detection concept is presented on Fig.5.

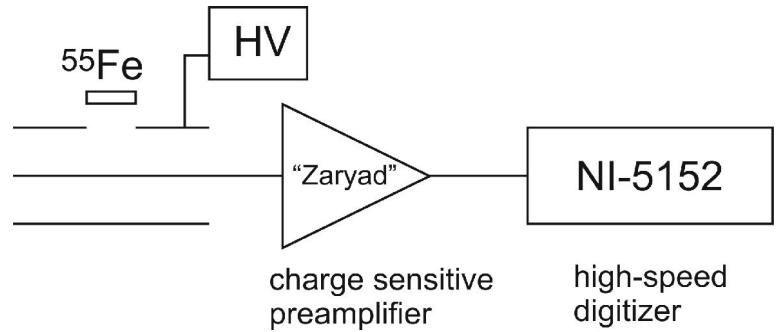

Figure 5. A schematic of the detection concept.

The calibration has been done using $^{55}Fe$ as a source of $X$-ray eradiation of 5.9 keV. The energy spectrum of the pulses from $^{55}Fe$ is presented on Fig. 6.

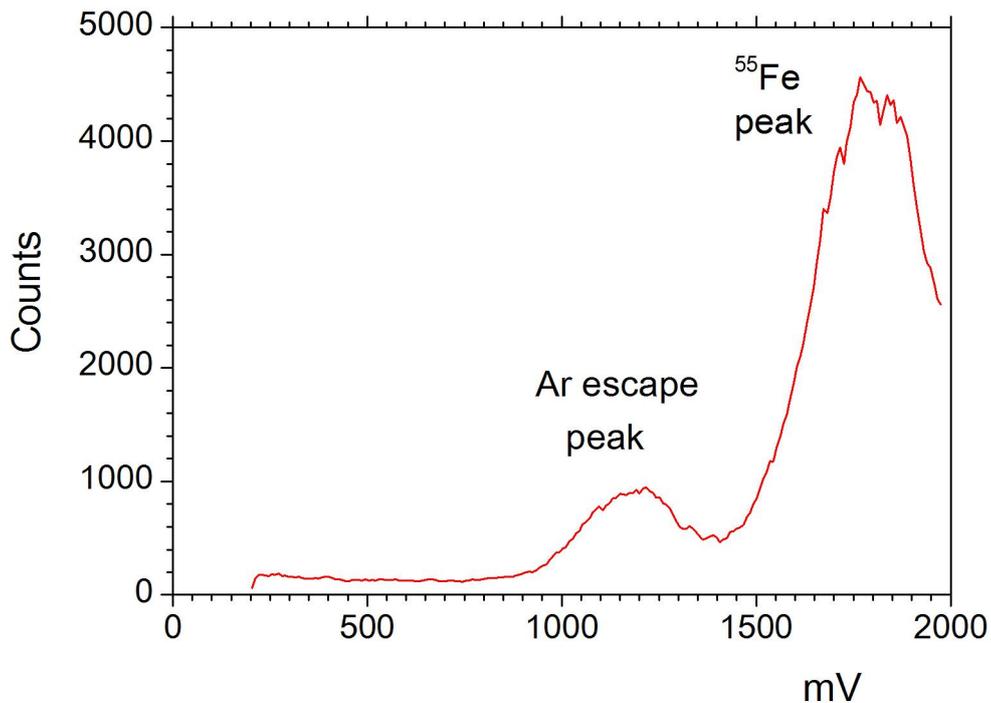

Figure 6. The energy spectrum from $^{55}Fe$ calibration source.

One can see two peaks: 5.9 keV from $^{55}Fe$ and 2.95 keV escape peak of argon. The resolution of the peak 5.9 keV of $^{55}Fe$ was 20% at 0.3 MPa of Ar + 10%CH$_4$ mixture. To get a high gas amplification of order $10^5$ we have increased High Voltage to 2760 V till the beginning of a regime of limited proportionality. Two peaks from $^{55}Fe$ source can be used for identification of the resulting non linearity. One can see this on Fig.7 where the amplitude in mV is presented as a function of the energy of the peak in keV.

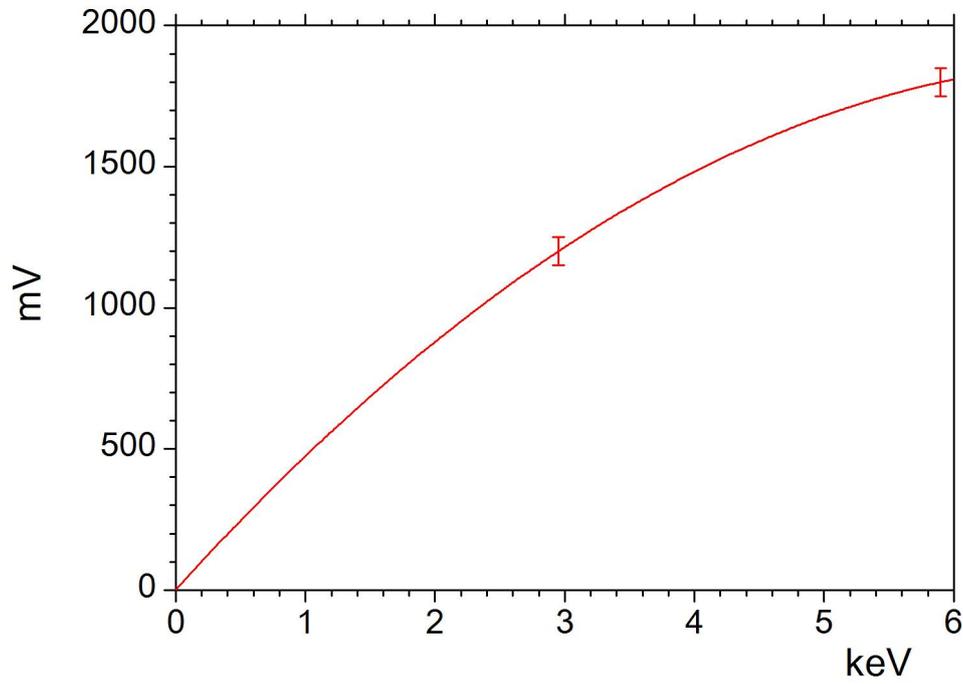

Figure 7. An observed amplitude of the pulse as a function of energy.

The shapes of the pulses from output of charge sensitive preamplifier of the sensitivity of about 0.5 V/pC have been recorded by 8-bit digitizer. The shapes recorded during certain time were analysed in off-line. The aim was to see how efficient could be the pulse shape discrimination of the noise pulses from electromagnetic disturbances and microphonic effect in the region below 100 eV, i.e. where the main effect is expected from CNNS of reactor antineutrinos. In Fig.8 we show the pulses observed during time interval 400 μs. For small amplitudes less than 100 mV the conversion factor was 0.6 mV/eV, see Fig.7.

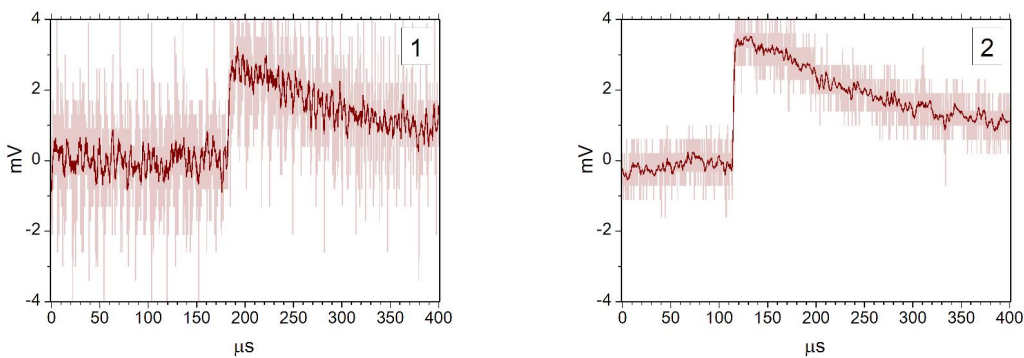

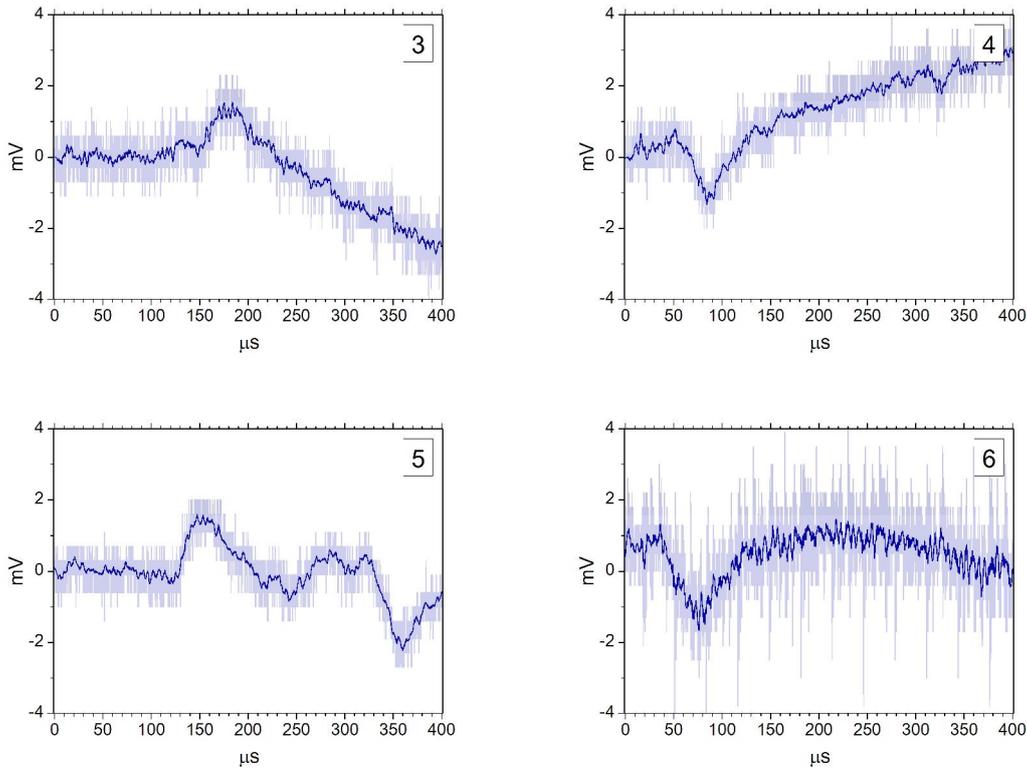

Figure 8. The shapes of the pulses in the time interval 400 μs.

Two snapshots, #1 and #2, are "true" pulses with correct signature from ionization process. One can see from these two pictures that the noise level somewhat varies in time and this influences the shapes of the pulses but still it is possible reliably select the pulses with amplitudes even as low as a few eV. The snapshots #3, #4, #5 and #6 show the "bad" pulses which are due to electromagnetic disturbances and microphonics. One can see the difference just by naked eye. The pulses from the point ionization in gas have typically a relatively short front edge (a few microseconds) corresponding to the time drift of positive ions to cathode and long (hundreds of microseconds) tail corresponding to the time of the base line restoration of the charge sensitive preamplifier. These events might be produced by internal radioactivity of the materials of the counter, by electronic emission from the walls of the counter and also by ionizing particles produced by cosmic rays. The amplitude of these events may be even smaller then an average energy to produce a single electron pulse because of the relatively broad energy distribution in this case (Polya distribution). In the range from 5 eV to 100 eV, where main effect from coherent scattering of reactor antineutrinos should be observed, the pulse shape discrimination enabled to reduce the noise by a factor of about $10^3$. The lower cut of 5 eV corresponds to the loss of efficiency of the detection of single electron events by approximately 5% which is quite appropriate for this experiment. Thus we show that this range can be effectively used for counting of the events from CNNS. The similar problem of counting the events from very small energy release has been solved in a number of experiments with cryogenic detectors. In 1997 we together with the staff of the laboratory of Professor S.Vitale in University of Genoa in Italy were first who succeeded in counting the pulses from peaks 57 eV and 112 eV from the decay of $^{7}Be$ [10]. The energy threshold in this work was 40 eV. This was achieved thanks to effective pulse shape discrimination of the noise pulses from electromagnetic disturbances and "microphonic effect".

3. The expected uncertainty of measurements.

The rate of counting for different gases for antineutrinos from reactor depends on the threshold of counting. The background count rate depends on the counting interval: the less is an interval, the less is the background. The optimal counting interval can be found from the calculated integral efficiency of counting presented on Fig.9.

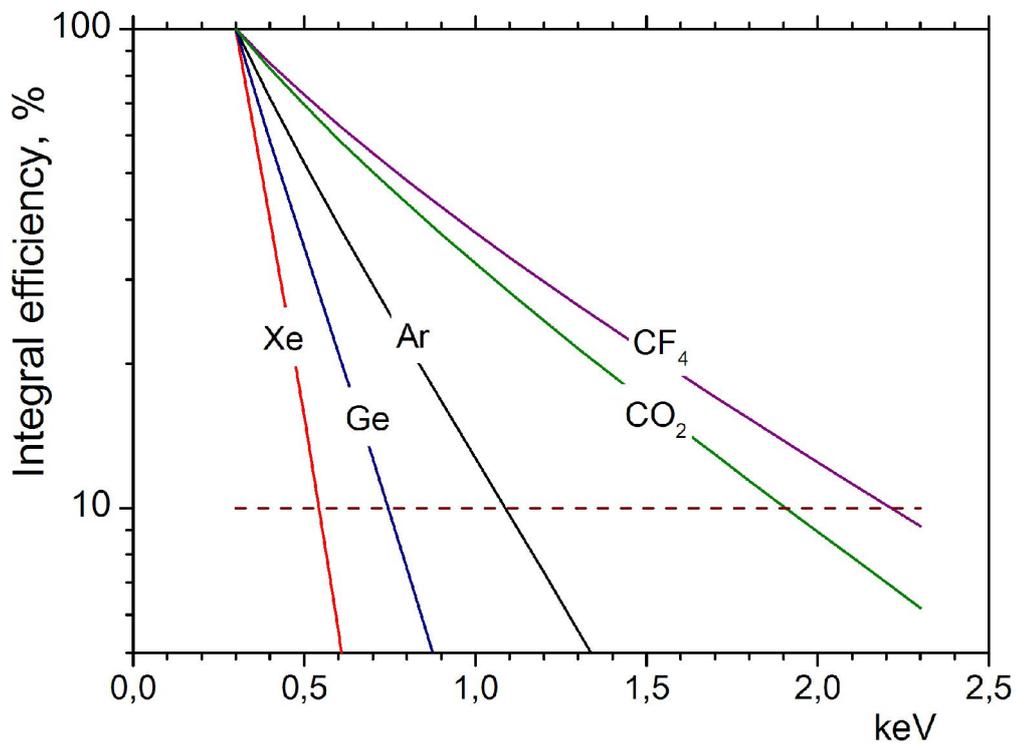

Figure 9. The integral efficiency of counting as a function of a threshold.

The crossing of the dashed line with curves for some specific gas indicates the interval which contains $\approx$ 90% of the total count rate higher than a threshold 300 eV at $QF$=1. From this point of view the most advantages would be to use xenon and the least advantages would be to use $CO_2$. In the real experiment $QF$ is expected to be on the level of 0.1 thus the real threshold will be around $\approx 30$ eV i.e. equal to an average energy of ionizing eradiation to produce an electron-ion pair in argon. However, $QF$ may be very different for different gases and for low nuclear recoil energies less than 1 keV there's no valid approximation which value of $QF$ to use. The measurement of this factor is very challenging and important task, but it is out of the scope of this paper. Here we consider two cases: when the background $B$ is equal to the rate $R$ and when it 10 times higher: $B = 10R$. We take a relative uncertainty $\delta$ equal to 10%, 5% and 2%. The first one is what we need to prove that we really observe the effect. The last one is what is necessary for the precise measurements to search for new physics. Having these $R$ and $X$ one write for the time of measurements needed to achieve these uncertainties as $t_1 = \dfrac{3}{R \cdot (0.01\delta)^2}$ for the first case and $t_1 = \dfrac{21}{R \cdot (0.01\delta)^2}$ for the second case. Table 1 shows the times needed for both cases for different gases. The flux of antineutrinos was taken $2.7 \cdot 10^{13}$ n/cm$^2$/sec, i.e. equal to the one at the site of KNPP in GEMMA experiment. One can see that for the vast majority of cases these times are quite reasonable and can be realized.

Table 1. The times of measurements in years at reactor on and reactor off to get a relative uncertainty $\delta$ at $B = R$ and $B = 10R$.

| Gas | $B = R$ | | | $B = 10R$ | | |
|---|---|---|---|---|---|---|
| | $\delta = 10\%$ | $\delta = 5\%$ | $\delta = 2\%$ | $\delta = 10\%$ | $\delta = 5\%$ | $\delta = 2\%$ |
| $^{40}$Ar | 0,037 | 0,141 | 0,889 | 0,248 | 0,996 | 6,219 |
| $^{132}$Xe | 0,019 | 0,081 | 0,507 | 0,141 | 0,567 | 3,541 |
| $^{74}$Ge | 0,022 | 0,085 | 0,533 | 0,148 | 0,596 | 3,726 |
| $CO_2$ | 0,052 | 0,215 | 1,341 | 0,374 | 1,504 | 9,396 |
| $CF_4$ | 0,022 | 0,089 | 0,548 | 0,152 | 0,615 | 3,841 |

The general conclusion from the data presented in Table 1 is that for 16 counters each containing 1 mol of gas the tolerable background is somewhere in between R and 10R. The notable point is that if we take argon from atmosphere the background from $^{39}Ar$ will be just in between these marks if to take the results of measurements presented in Ref [11]. The use of argon from underground sources depleted by $^{39}$Ar may turn out to be a very useful approach. The background from neutrons generated in the surrounding materials by cosmic rays was calculated as it was described in Ref. [12]. It is presented in Fig.2 for different gases and also for germanium just for comparison. One can see that it is quite tolerable to conduct the experiment and it can be efficiently suppressed by veto of an active shield.

**Results and Discussions.**

The measurements were performed on argon proportional counter at pressure from 0.1 till 0.3 MPa using $^{55}Fe$ calibration sources to record the shapes of the pulses from ionization with amplitudes from a few eV till a few keV. It was shown that the use of gaseous proportional counters with gas amplification of about $10^5$ and low noise electronics enable the reliable registration of the ionizing eradiation with the energy threshold as low as 5 eV. The pulse shape discrimination in the energy interval from 5 eV till 100 eV enables substantially reduce noise from electromagnetic interferences and microphonics. The critical point to conduct the experiment is the background. The one from neutrons generated in the surrounding materials by cosmic rays was calculated for different working mediums. The detector should be placed as close as possible to the core of the reactor to get high flux of antineutrinos. So the overburden in the experiment is determined by the construction of the reactor and cannot be changed. The calculation shows, see Fig. 2 that the background level from neutrons is quite satisfactory and can be tolerated. The total background from all sources, internal and external can be tolerated if it is somewhere in between two marks: $B = R$ and $B = 10R$ and it determines the time of measurements needed to get the definite uncertainty. One can see from Table 1 that it is quite appropriate from experimental point of view. However it demands that a very clean materials be used for the fabrication of the detector and in case of argon it would be helpful to use an underground argon depleted by $^{39}Ar$. The important point for verification of the discovery of CNNS would be to make experiment with different gases, say, argon and xenon to compare the results obtained and to prove that the spectral data are in agreement with the predicted ones for CNNS.

**Conclusions.**

The gaseous proportional counter with a threshold 5 eV is a very perspective tool to study neutrino scattering on electrons and nuclei at small recoil energies. It has a potential of making discovery of still unobserved effect from coherent scattering of neutrinos on nuclei and may measure the magnetic moment of neutrinos, if it is higher than $5 \cdot 10^{-12}$ μB. To have a mass adequate to obtain the accuracy of measurements of a few percent the detector should be composed of several counters. We are planning to use an array of 16 counters arranged in 4 planes. The main limiting factor is the background so the detector should be constructed from very pure materials, well shielded from external radiation and neutrons and equipped with an electronics system with pulse shape discrimination. At the present time the work if focused on the construction of one prototype module and R&D on this module. The next step will be the construction of the array from 16 modules and placing the full assembly with active and passive shielding at the reactor. This work was partially supported by a grant of Leading Scientific Schools of Russia LSS #3110.2014.2.